\documentclass[aps,preprintnumbers,superscriptaddress,twocolumn,nofootinbib]{revtex4}
\usepackage{epsfig,graphics}
\usepackage{amsfonts,amssymb,amsmath}            
\usepackage{amsthm}
\usepackage{multirow}                           
\def\identity{\openone}
\def\half{\frac{1}{2}}

\newcommand{\ket}[1]{\left |  #1 \right\rangle}

\newcommand{\comment}[1]{}
\theoremstyle{plain}
\newtheorem{theorem}{Theorem}
\newtheorem{lemma}{Lemma}
\newtheorem{corollary}{Corollary}

\theoremstyle{definition}
\newtheorem{definition}{Definition}

\begin{document}

\title{Quantum Self-Correcting Stabilizer Codes}
\date{\today}

\author{Alastair \surname{Kay}}
\email[]{a.s.kay@damtp.cam.ac.uk}
\affiliation{Max-Planck-Institut f\"ur Quantenoptik, Hans-Kopfermann-Str.\ 1,
D-85748 Garching, Germany.}
\affiliation{Centre for Quantum Computation, Department of Applied
Mathematics and Theoretical Physics, University of Cambridge, Cambridge CB3
0WA, UK.}
\author{Roger \surname{Colbeck}}
\email[]{colbeck@phys.ethz.ch}
\affiliation{Institute for Theoretical Physics, ETH Zurich, 8093
  Zurich, Switzerland.}
\affiliation{Institute of Theoretical Computer Science, ETH Zurich, 8092
  Zurich, Switzerland.}

\begin{abstract}
  In this paper, we explicitly construct (Abelian) anyonic excitations
  of arbitrary stabilizer Hamiltonians which are local on a 2D lattice
  of qubits. This leads directly to the conclusion that, in the
  presence of local thermal noise, such systems cannot be used for the
  fault-tolerant storage of quantum information by self-correction
  i.e.~they are ruled out as candidates for a `quantum hard drive'.
  We suggest that in 3D, the same construction leads to an argument that self-correction is impossible.
\end{abstract}

\maketitle

\section{Introduction}
Our capacity for exploiting the properties of quantum systems for
information processing tasks is critically dependent on the ability to
protect this fragile information against the unwanted destructive
effects of the environment. While theories of error correction
\cite{Ste96c,CS96a,Got98b} and fault-tolerance
\cite{Ste98a,Shor_faults,aharanov:99,gottesman:2005,Kni04a,AK3} have
been proposed, these might be considered early steps towards proving
what is possible, but that the resource requirements are prohibitive
for useful implementation. A possible candidate for a `second
generation' architecture for fault-tolerant storage of information is
known as self-correction. This concept has arisen from the study of
the toric code, where information is encoded in the degenerate ground
states of a Hamiltonian. If the energy cost for converting between the
degenerate states using local operations grows with the number of
qubits in the lattice, then logical errors can potentially be
exponentially suppressed as the system size scales. This property is
known as self-correction, and, if systems with such properties could
be found, may reduce the energy cost of information storage. To date,
no proof of self-correction exists, although candidate models in three
\cite{Bac05a} and four \cite{DKLP02a} spatial dimensions have been
proposed.

In this paper, we investigate which Hamiltonians can possibly present
self-correcting properties. Previous studies have restricted
themselves to the toric code
\cite{horodecki,nussinov-2007,nussinov-2007a,CC07,Kay:08,alicki},
except for one attempt at a more general no-go theorem
\cite{horodecki}. However, this latter result seems to address the
question of whether one can store information in a thermal state
rather than how long it takes for a system to thermalize.

Here we prove that stabilizer Hamiltonians on qubit lattices in two
spatial dimensions are not self-correcting i.e.\ the time required for
the system to develop an error via a local noise model is not
exponentially increased by enlarging the lattice.  Our proof
involves explicitly constructing paths through which such noise
destroys the stored information. We also argue the existence of such
paths in 3D.

Let us begin by precisely defining what we are interested in, and the
goal of this paper.
\begin{definition}
  A topological quantum stabilizer code in two spatial dimensions is
  an instance from a family of qubit Hamiltonians parameterized by
  their size ($N$). The Hamiltonian is a sum of $R$ terms $K_n$, where
  $[K_n,K_m]=0$ and $K_n^2=\identity$.  Each $K_n$ is a tensor product
  of operators confined to a local area on a 2D lattice, and whose
  size is independent of $N$.
\end{definition}
Such codes can be designed such that there are degenerate ground
states of the Hamiltonian in which quantum information can be stored.
The classic example is the toric code \cite{DKLP02a}.  For
concreteness, we restrict to the case of a square lattice with
periodic boundary conditions, although the same arguments can be made
for any planar lattice geometry and, with some modification, open boundary conditions.

There are many reasons why restricting to stabilizer codes is
beneficial. The mutual commutation of terms makes calculations simpler
than for general Hamiltonians, and thus provides a natural starting
point. In particular, the ground state degeneracy is readily
identified (and, usefully, the degenerate subspace persists at all
energy levels). It also means that the implementation of error
detection is easily described. Even in a self-correcting system, error
detection and correction is important|at some stage it will be
necessary to read out the stored information, and, while self
correction will ensure that the number of errors is small and that
consequently there will be no false read-out of the state, one
nevertheless needs to detect the errors that have occurred.

Furthermore, one might hope that by examining the case of
stabilizer codes, these turn out to be representative of a large class
of Hamiltonians. For example, in the 1D scenario, there are some
generic properties of cluster states (which are a specific example of
a stabilizer state), such as correlation functions, which, when looked
at on a global scale, are very close to those for all local 1D
gapped Hamiltonians.  Specifically, two-body correlations of the
ground states of these systems decay exponentially with distance
\cite{hastings1,hastings2}, and are hence negligibly small when we
examine them over separations that scale with system size. In
comparison, the correlation functions of the cluster state are
identically zero \cite{vbs:1}. Adding rigour to this possible
connection is beyond the scope of the present work, but is certainly
an interesting avenue for future studies.  Nevertheless, this may be
considered a difficult proposition.  Properties of the ground states
are certainly going to involve entanglement in some way, and yet our
understanding of multipartite entanglement is quite poor. For
instance, while we are able to discuss distillation of noisy
stabilizer states \cite{purification:1,purification:2}, including some
optimality results \cite{AK1,AK2}, results relating to non-stabilizer
states are virtually non-existent \cite{miyake}, and have only been
realized by mapping the system into a stabilizer form (with the
exception of some special cases developed in \cite{AK2}).

The restriction to qubits is, again, a matter of convenience since it
imposes a number of useful properties on the stabilizers. Furthermore,
the literature on binary quantum codes, which have a direct relation
to stabilizer states \cite{schlingemann,Schlingemann01}, is vastly
more developed than the non-binary case \cite{rains,knill:01}. As a
result, the generalization to qudits is not immediate.

Topological quantum memories are a useful step towards building
fault-tolerant quantum computers, giving a degree of protection
against Hamiltonian perturbations. However, as will be discussed in
Sec.~\ref{sec:1}, local thermal noise can destroy the stored
information, in spite of the apparent protection due to the existence
of an energy gap
\cite{DKLP02a,horodecki,nussinov-2007,nussinov-2007a,CC07}. This means
that active error correction is required, involving a supply of fresh
ancillas, or a dissipative operation to reset them. Since the toric
code is a stabilizer code, the error syndrome can be extracted by
measuring the stabilizer operators, and hence the physical action of
correction is relatively simple.  However, it is more desirable to
have self-correcting codes; those that exponentially suppress errors
due to both Hamiltonian perturbations and thermal noise, without
needing external interaction.  One could simply leave such a memory in
a `power off' state and expect it to remain in the meta-stable state
for a time that scales exponentially with the size of the system.
Such a system could form a `quantum hard drive' which would enable
high fidelity storage of quantum information, as well as forming a
building block for fault-tolerant information processing and
Hamiltonian simulation.

\begin{definition}
  A quantum self-correcting stabilizer code is a topological quantum
  stabilizer code, with the additional property that
  the survival time of the quantum information in the presence of
  local thermal noise, without active error correction except at the
  final, read-out, phase, scales exponentially with $N$.
\end{definition}
While we cannot expect to store information for an arbitrarily long
time, such a definition allows a beneficial scaling of the protection
against errors with the size of the system. Our central thesis is that
there are no quantum self-correcting stabilizer codes in two spatial
dimensions.  However, we have not proven this in the full generality,
rather, our proof only applies to \emph{specified} systems, where all
the degeneracy is caused by products of stabilizers being identity.  A
trivial example where this is not the case would be defining an
$N$-qubit Hamiltonian but considering $N+1$ qubits|this would be
under-specified on the additional qubit.

The statement of our theorem is
\begin{theorem}
\label{thm:main}
There are no quantum self-correcting stabilizer codes in two spatial
dimensions in the case of \emph{specified} Hamiltonians.
\end{theorem}

The proof of this will appear in Sec.~\ref{sec:proof}.

\section{Preliminaries} \label{sec:1}

In this section we illustrate the use of stabilizer systems for
storing quantum information with two well-known examples, the 2D Ising
model and the toric code. The 2D Ising model provides robust storage
of classical information, while quantum information can be destroyed
by a single local operation, whereas the toric code puts the logical
$X$ and $Z$ rotations of a qubit on an equal footing, and we will
discuss why the existence of these string-like operators implies that
the model is not self correcting. This motivates our study, where we
show that all stabilizer Hamiltonians in 2D behave like one of these
two models, so there is no quantum self-correction.

\subsection{Classical Self-Correction and the Ising Model}

While we have defined self-correction and topological memories for the
storage of quantum information, identical concepts exist in the
classical case. In order to understand why systems such as the toric
code are not self-correcting, it is instructive to
examine the classical case, specifically the Ising model. In one
dimension, the Ising model is a classical memory, but is unstable
against thermal noise, whereas the Ising model in 2D is a
self-correcting code for classical information. It is already known
that the string-like properties of the toric code can be transformed into the 1D Ising model (in
fact, two parallel copies) \cite{nussinov-2007,nussinov-2007a}.

The one-dimensional Ising model,
$$
H=-\frac{1}{2}\Delta\sum_iZ_iZ_{i+1}
$$
has degenerate ground states $\ket{00\ldots 0}$ and $\ket{11\ldots 1}$
which can encode a bit, and a gap to the first excited state of
$2\Delta$. A qubit cannot be reliably encoded because a single $Z$
rotation has no energy cost, and performs a logical $Z$-rotation on
our bit, so there is no protection against this type of error. In
contrast, to perform a bit-flip (logical $X$) one must apply the
operator $X^{\otimes N}$. In the presence of noise, this can be realized with a sequence of local $X$ operations, the first of which costs an energy $2\Delta$ (which is recovered at the last step), with the remainder having no energy cost. Thus, while there is some protection afforded by the system against the first error, subsequent errors are not prevented, and one can arrive very quickly at a logical rotation.

In comparison, the two-dimensional Ising model on an $N\times N$
lattice has the Hamiltonian
$$
H=-\frac{1}{2}\Delta\sum_{\langle i,j\rangle}Z_iZ_j,
$$
where $\langle i,j\rangle$ denote neighbouring lattice sites.  Like
its one-dimensional counterpart, it has the degenerate ground states
$\ket{00\ldots 0}$ and $\ket{11\ldots 1}$, so again only permits
a classical code.  However, in this case, a single bit flip, which
costs $4\Delta$, is locked|to flip another neighbouring spin has a
further energy cost of $2\Delta$. The energy cost of flipping a block
grows with the surface area of the block, and so it becomes extremely difficult for an environment to perform a logical $X$ operation. These concepts have been made rigorous in a number of ways. For example, it was shown in \cite{Kay:08} how perturbations on the Hamiltonian can propagate single errors into logical errors. For the 1D Ising model, the toric code, and indeed any stabilizer model with string-like logical operations, the time required is polynomial in the system size. A fixed density of errors can be converted into a logical gate operation in a time independent of the lattice size. On the other hand, it was also shown that such conversions in the case of the 2D Ising model require an exponentially long time, precisely because of the energy structure occurring due to the two-dimensional topology of the logical gate operation.

\subsection{The Toric Code}

\begin{figure}
\begin{center}
\includegraphics[width=0.3\textwidth]{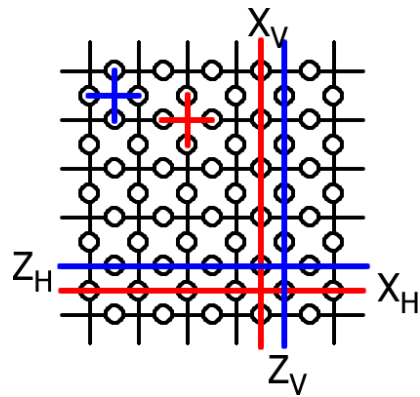}
\end{center}
\caption{The toric code is defined on a square lattice as a sum of
  4-body terms (qubits are indicated by circles). On the faces of the
  squares, there are terms $XXXX$ (red cross) and at the corners
  there are terms $ZZZZ$ (blue cross). There are 4 string operators
  making non-trivial loops around the system, composed of single rows
  or columns of Pauli operators $X$ and $Z$.} \label{fig:toric}
\end{figure}

The toric code, as depicted in Fig.~\ref{fig:toric}, is described by a
Hamiltonian
$$
H=-\half\Delta\left(\sum_S XXXX+\sum_P ZZZZ\right),
$$
where the sums are taken over the plaquettes, $P$, (the sets of four
qubits which surround a square) and sites, $S$, (the sets of four
qubits which surround a corner). This model can be transformed into
two parallel copies of the one-dimensional Ising model, and it is this
transformation which indicates that the toric code is not
self-correcting. Specifically, to perform operations within the ground
state space of the Hamiltonian, strings of $X$ and $Z$ operations
around the two topologically inequivalent loops of the torus are used.
These are denoted $X_H$ and $X_V$ for the `horizontal' and `vertical'
loops respectively. One potentially useful property of the toric code is that if the Hamiltonian is affected by
a sum of local perturbations, $V$ ($\|V\|\ll1$) , one has to consider $N^{{\rm
th}}$ order perturbation theory before the errors can possibly compose themselves into a string that affects ground state space. Since these are of strength $\|V\|^N$, a linear increase
in the size of the lattice yields an exponential suppression in error.

However, if an environment is able to apply local rotations, the sequential
flipping of the spins allows a logical rotation to be implemented.  Such flips have the same energies as those of the
bit-flips in the 1D Ising model, i.e.~once the initial excitation has
been created (with a single-qubit rotation), there is no further
energy cost in extending the string.

For stabilizer codes, there is a concrete relation between the
operations that convert between the degenerate ground states and the
error pathways. Assuming that such an operation constitutes a
string-like tensor product of Pauli operators around a loop, a
truncation of this loop only has a finite energy because only
stabilizers overlapping with the ends can possibly anti-commute with
it. These string-like loops, if they exist, therefore describe a path
through which noise destroys the stored information. This is exactly
what is required for the application of the result in \cite{Kay:08}.
Other studies such as \cite{alicki} reveal a similar result.

\section{Stabilizer Hamiltonians and Degeneracies}

Consider a set of $R$ stabilizer operators $W_H=\{K_n\}$,
$
[K_n,K_m]=0,
$
which are used to construct a Hamiltonian
$$
H=-\half\Delta\sum_{K\in W_H}K
$$
applied on an $N\times N$ square qubit lattice with periodic boundary
conditions. The $\{K_n\}$ are tensor products of Pauli operators
acting on systems of qubits, and hence have eigenvalues $\pm 1$. Since
any tensor product of Pauli operators has trace $0$ (except the
trivial case of $\openone$), exactly half of the eigenvalues are $1$
and half are $-1$. We shall take each of the $K_n$ to be non-identity
on at most a $k\times k$ block of spins on the 2D lattice, where $k$
is some fixed number, and $N\geq 2k$.  In general, there may be values
of $n$ for which $K_n$ acts on a smaller area.

It is convenient to place the stabilizers in sets, such that the
product of all stabilizers in each set is $\identity$.  The
combination of two such sets generates a third set which also product
to $\identity$, where sets are combined under the operation
$(G_i,G_j)\equiv G_i\cup G_j\setminus G_i\cap G_j$, with
$(G_i,G_j,G_k)\equiv(G_i,(G_j,G_k))$ and so on.
This motivates the following definitions.
\begin{definition}
  Let $\{G_i\}$ be \emph{identity sets} and denote $G$ as the set of all
  such sets.  Then, for all $G_i\in G$,
\begin{equation}
  \label{eqn:degen}
  \prod_{K\in G_i}K=\identity.
\end{equation}
\end{definition}

\begin{definition}
Let $G'_i$ be a \emph{minimal identity set} and denote
  $G'$ as the set of all such sets.  Then, for all $G'_i\in
  G'$,
\begin{equation*}
  \prod_{K\in G'_i}K=\identity.
\end{equation*}
holds, but is not implied by $\left\{\prod_{K\in G'_{k\neq
  i}}K=\identity\right\}$.
\comment{Further, there is no subset
  $W\subset G_i\in G$ for which $\prod_{K_j\in W}K_j=\identity$.}
\end{definition}

Note that the identity sets can be formed from the minimal set of
identity sets by combining them under the operation
$(G_i,G_j)$.  The number of combinations is
\begin{equation}
|G|=\sum_{i=1}^{|G'|}\,^{|G'|}\text{C}_i=2^{|G'|}-1.
\end{equation}

\begin{lemma}
  Consider a Hamiltonian, $H$, composed of $R$ stabilizers defined on
  an $N\times N$ square lattice of qubits. Such a
  Hamiltonian has a ground state degeneracy of $2^M$ levels, where
$$
2^M=2^{N^2-R}(1+|G|).
$$
\end{lemma}
\begin{proof}
  The ground state space of $H$ is given by projection onto the $+1$
  eigenstate of each of the stabilizers,
\begin{equation}
\label{eqn:rho}
\rho=\frac{\prod_n(\identity+K_n)}{\text{Tr}\left(\prod_n(\identity+K_n)\right)}
\end{equation}
and the degeneracy is given by $2^M=\text{rank}(\rho)$. Alternatively,
we can calculate the degeneracy with
\begin{eqnarray}
2^M&=&\frac{1}{2^R}\text{Tr}\left(\prod_n(\identity+K_n)\right)\\
&=&\frac{1}{2^R}\text{Tr}\left(\identity+\sum_iK_i+\sum_{i,j>i}K_iK_j+\ldots\right).\label{eqn:5}
\end{eqnarray}
The expression within the trace is composed of ordered products of the
stabilizers, and the trace is hence equal to the number of times the
stabilizers product to $\identity$, multiplied by a factor of
$\text{Tr}(\identity)=2^{N^2}$. All sets that product to identity
contribute to the sum, and there is an additional contribution from
the first term in the trace in (\ref{eqn:5}), hence
$$
2^M=2^{N^2-R}(1+|G|).
$$
\end{proof}

Note that there are two possible causes of degeneracy: the first is
that $R<N^2$, which means that there are insufficient stabilizers to
break all the potential degeneracies.  The second is that $|G|>0$,
i.e.\ products of stabilizers are identity.  We show in this paper how
to break degeneracies of the second type.  A system is said to be
\emph{specified} if it only has degeneracies of this type.

\begin{definition}
  Two operators, $S_1$ and $S_2$ are said to be \emph{independent} with
  respect to a set of stabilizers $G_i$ if there is no set $W\subseteq
  G_i$ for which $S_1S_2=\prod_{K\in W}K$, i.e.\ if they are not
  related by a product of stabilizers in $G_i$.  Otherwise, they are
  \emph{dependent} with respect to $G_i$. If no set $G_i$ is indicated,
  the entire set of operators is intended.
\end{definition}

For the sake of the general proof, it is helpful to define a subset of
the minimal identity sets $G'$, which we refer to as the elementary sets, $\tilde G$. These are closely related to the concept
of dependence of operators.
\begin{definition}
  Consider a subset of the $R$ stabilizers, $W_R$, where
  $|W_R|=N^2-M$, such that these stabilizers contain all the
  information of the Hamiltonian (i.e.~$W_R$ does not contain any identity sets, and thus all stabilizers not in $W_R$ can be
  generated by products of members of $W_R$). The elementary sets $\tilde G_i$ are the
  minimal identity sets formed by adding back $|\tilde G|=M$
  stabilizers such that each $\tilde G_i$ corresponds to a degeneracy.
\end{definition}

\begin{definition}
  We say an identity set $G$ is \emph{topologically trivial} in the
  vertical direction if there exists a row, $t$, such that after removing all
  stabilizers which have support both above and in row $t$ from $G$,
  the remaining stabilizers still product to identity in and
  below row $t$.  Otherwise, the set $G$ is \emph{topologically
    non-trivial} \footnote{In the case of open boundary conditions, the definition of an identity set must be changed, specifically in the case of topologically non-trivial identity sets. By assuming periodic boundary conditions, while allowing arbitrary spatial variation of the stabilizers, we are assuming that this situation does not arise.}.
\end{definition}

There is much freedom in the choice of $W_R$, and also in the choice
of stabilizer to add in to form the elementary sets.  We will assume,
without loss of generality, that the elementary sets are picked to be
topologically non-trivial as far as possible, i.e.\ they consist of
stabilizers that extend around the entire lattice in some direction.
We will also show (in Lemma \ref{lemma:7}) that we can choose any
topologically trivial elementary sets to be local (i.e.\ containing
stabilizers only in a $k\times k$ area).

Since we are assuming that the Hamiltonian is specified, we have that all the
stabilizers $W_R\in W_H$ and $\tilde G_i\in W_H$ are already contained in the
Hamiltonian, and hence are local.

\section{Constructing String Operators for Stabilizer Codes}

\subsection{Proof Sketch}

In order to show that a stabilizer Hamiltonian cannot store quantum bits in
a self-correcting way, we show that there exists a product of local operators which
converts between each of the ground states, the spatial patterning of which forms a one-dimensional structure. Such an operator is called a \emph{string
operator}.  It
is this one dimensional nature that prevents the exponential
suppression of errors.  Since the entire operator commutes with all
the stabilizers, if one considers building it up by applying the local rotations sequentially, only the ends of the sequence anti-commute with terms in the Hamiltonian and hence the size of
these bound the energy of the partial string to a constant level. Any proof regarding the fragility of quantum information which is applicable to the toric code is equally applicable here.

Each ground state degeneracy can be attributed to the existence of a
set of stabilizers (a member of $\tilde G$) that product to identity.
Our proof will constitute taking each of these sets in turn and
constructing from them an operator that has the structure of a
one-dimensional loop around the lattice and commutes with all the
stabilizers of the system.  Such an operator thus provides a path for
errors to propagate. The critical component to this proof is to
realise how to construct an operator that automatically commutes with
all stabilizers|we do this by using the fact that the stabilizers
themselves already mutually commute, so products of them must also do
so.  Furthermore, to guarantee commutation of two stabilizers, one
does not need to retain the full stabilizer, only the areas that
overlap.

Once we have established that there are loop operators arising from
each of the sets in $\tilde G$, we prove that they are independent
i.e.~that error pathways from two different loops have different
effects on the degenerate space, so we are guaranteed that all the
degeneracies are broken.

Before embarking on the complete proof, whose essence may be obscured
by technicalities, we present the construction for the case of a
translationally invariant system where the stabilizers act on $3\times
3$ blocks of qubits.

\subsection{An Example: $3\times 3$ Translationally Invariant Codes}

The simple case of $k=3$, where all the stabilizers are the same (just
displaced to different sites), removes many of the technicalities
required for the general proof and, as such, provides a useful example
that is entirely consistent with what will be presented later. Due to
the translational invariance, it must be true that the product of all
the stabilizers is equal to $\identity$, in order for there to be some
degeneracy present.  From now on, we assume this to be the case.

For each lattice site, there are 9 stabilizers which overlap with it,
one for each term in the stabilizer. Since the product of stabilizers
on this site is $\identity$, we conclude that the product of all
operators in the stabilizer must be $\identity$. Let us represent the
stabilizers as follows:
\begin{equation}
\label{eqn:operator}
\begin{array}{ccc}C_{11}&C_{12}&C_{13}\\C_{21}&C_{22}&C_{23}\\C_{31}&C_{32}&C_{33}\end{array},
\end{equation}
where $\{C_{ij}\}$ are Pauli operators.  We denote products of all the
operators in a particular row of a stabilizer as $R_i:=\prod_jC_{ij}$.
Due to local unitary equivalence, without loss of generality, we take
the first non-$\identity$ operator to be $X$. Since
$R_1R_2R_3=\identity$, there are only a limited set of possibilities
that we have to consider:
\begin{equation}
(R_1,R_2,R_3)\in\{(\identity,\identity,\identity),(X,\identity,X),(X,X,\identity),(X,Y,Z)\}.	\label{eqn:row_opts}
\end{equation}
Necessary conditions for (\ref{eqn:operator}) to be a stabilizer (i.e.~for it to commute with itself in different positions on the lattice) are that
\begin{equation}
[R_1,R_3]=0\quad\text{and}\quad[R_1\otimes R_2,R_2\otimes R_3]=0.	\label{eqn:row_products}
\end{equation}
We proceed to prove $[R_1,R_3]=0$, since this is the only condition that we require. In order for the operator defined by (\ref{eqn:operator}) to be a
stabilizer, it must satisfy the following commutation relations,
\begin{eqnarray}
\left[C_{13}\otimes\identity\otimes\identity, C_{31}\otimes C_{32}\otimes C_{33}\right]&=&0	\nonumber\\
\left[C_{12}\otimes C_{13}\otimes\identity, C_{31}\otimes C_{32}\otimes C_{33}\right]&=&0	\nonumber\\
\left[C_{11}\otimes C_{12}\otimes C_{13},C_{31}\otimes C_{32}\otimes C_{33}\right]&=&0	\nonumber\\
\left[\identity\otimes C_{11}\otimes C_{12},C_{31}\otimes C_{32}\otimes C_{33}\right]&=&0	\nonumber\\
\left[\identity\otimes\identity\otimes C_{11},C_{31}\otimes C_{32}\otimes C_{33}\right]&=&0,	\nonumber
\end{eqnarray}
where we have retained unnecessary terms so that the following step
is clearer. Now we use the fact that if $[A,C]=0$ and $[B,C]=0$, then
$[AB,C]=0$, combining all 5 equations to give
\begin{equation}
\label{eqn:Rcom}
[R_1\otimes R_1\otimes R_1,C_{31}\otimes C_{32}\otimes C_{33}]=0.
\end{equation}
For Pauli operators $\{P_i\}$ and $P$,
\begin{equation}
\left[\prod_i^m P_i,P\right]=0\:\Leftrightarrow\:\left[P_1\otimes\ldots\otimes
P_m,P\otimes\ldots\otimes P\right]=0.
\end{equation}
To see this, note that the
  LHS and RHS both imply that an even number of $\{P_i\}$ anticommute
  with $P$.
Hence, applying this to (\ref{eqn:Rcom}) shows that $[R_1,R_3]=0$ is a
necessary condition for the operator to be a stabilizer.

This eliminates the $(X,Y,Z)$ possibility from
Eqn.~(\ref{eqn:row_opts}). Now consider the cases $(X,\identity,X)$
and $(X,X,\identity)$ in Eqn.~(\ref{eqn:row_opts}). One finds that the
string operators
\begin{equation*}
\begin{array}{ccccc}\multirow{2}{*}{\ldots}&R_1&R_1&R_1&\multirow{2}{*}{\ldots}\smallskip\\&R_1R_2&R_1R_2&R_1R_2&\end{array}
\end{equation*}
commute with the stabilizers, i.e., we get either two parallel rows of
$X$s, or a single row of $X$s. One can verify that in both cases, a
single row of $X$s commutes with the stabilizers by using
(\ref{eqn:Rcom}). In fact, the case of $(R_1,R_2,R_3)=(X,\identity,X)$
has degeneracy of at least $2^M=4$, since we can take the product of
all stabilizers on every second row and get $\identity$. As will
become clear in the general proof, this accounts for the difference
between taking a single row of $X$s and two rows; a single row of $X$s
on row 1 compared to a single row of $X$s on row 2 will implement a
different logical operation on the logical qubits.

The remaining case is $(R_1,R_2,R_3)=(\identity,\identity,\identity)$.
This case has many degeneracies since a single row of stabilizers
products to $\identity$, and there are $N$ such rows. To deal with
this case, one can first analyse the analogous column operators
constructed for the row of stabilizers that product to identity.  An
identical construction can be performed on these, except that the
resulting operators only extend over 3 qubits, the height of the row,
rather than taking the form of loops.  Such operators are readily
implemented by local noise.  However, this construction also fails if
the columns product to $(\identity,\identity,\identity)$ as well.  In
such a case, note that each column of the stabilizer must be either
$(\identity,\identity,\identity)$, $(X,X,\identity)$ or $(X,Y,Z)$ (up
to permutations).  Any column operator $X^{\otimes 3}$, $Y^{\otimes
  3}$ or $Z^{\otimes 3}$ commutes with all of these, and therefore
commutes with all the stabilizers.  Such an operator splits the ground
state degeneracy, and is a fixed size, hence the system is again
susceptible to local noise and cannot give any useful protection.

\subsection{String Operators}

\begin{definition}
  A \emph{string operator} $S$ is a tensor product of localized
  operators which commutes with the stabilizers, $[S,K_i]=0$, has
  eigenvalues $\pm 1$ and is defined over an area of the lattice which
  scales with $N$ in no more than one dimension.
\end{definition}
For example, the string operator for the one dimensional Ising model
is $X^{\otimes N}$. In comparison, the required operator to
inter-convert ground states for the 2D Ising model extends over two
dimensions, and is therefore not a string operator. The single $Z$
rotations that apply logical phase gates are also string operators in
both cases, although we may choose to refer to them as point operators
since they have no dependence on $N$. Our aim is to find $|\tilde G|$ string
operators $\{S_i\}$ such that each $S_i$ performs an independent
non-trivial action on the degenerate ground states.  Since each $S_i$
is a tensor product of Pauli operators and identity, half of the
eigenvalues of $S_i$ are $+1$ and half are $-1$.  A string operator
with \emph{non-trivial} action splits the degenerate ground space in
half.
\begin{lemma}
  An operator, $S_i$, which satisfies $[S_i,K_n]=0$ for all $n$, performs
  a non-trivial action within the ground state space of $H$ provided
  that $S_i$ is independent of $\identity$.
\comment{there is no set, $W_i$, for which $S_i=\prod_{K_n\in W_i}K_n$.}
  \label{lem:action}
\end{lemma}
\begin{proof}
  Consider the projection of $\rho$ (as defined by (\ref{eqn:rho}))
  onto the $+1$ eigenspace of $S_i$,
$$
\rho'=\half(\identity+S_i)\rho.
$$
If there exists a set $W$ such that $S_i=\prod_{K\in W}K$, then we
already know that $S_i$ must have eigenvalue $+1$ when acting on
$\rho$, hence $S_i$ has trivial action.

Conversely, if there is no set $W$, then
$\text{Tr}(\rho')=\half\text{Tr}(\prod_j(\identity+K_j))=2^{M-1}$, so
the ground state space is split in half, i.e.~within the space defined
by $\rho$, there are states $\ket{\psi_{\pm}}$ which satisfy
$S_i\ket{\psi_{\pm}}=\pm\ket{\psi_{\pm}}$ so $S_i$ acts like a logical
phase gate on this space. (Alternatively, we could rewrite it to
appear as a logical $X$ rotation, using basis states
$(\ket{\psi_+}\pm\ket{\psi_-})/\sqrt{2}$.)
\end{proof}
\begin{lemma}
\label{lem:sameact}
  Two dependent operators $S_i$ and $S_j$ have the same action on the
  ground states.
\end{lemma}
\begin{proof}
  The state $\rho'$ has eigenvalues $+1$ for all $K$ and $S_i$.
  Given that $S_j$ is dependent on these, there exists a $W$ such that
  $S_j=S_i\prod_{K\in W}K$, and hence the value of $S_j$ when
  acting on $\rho'$ is $+1$, and the action of $S_j$ is trivial
  i.e.~it had the same effect as $S_i$.
\end{proof}

Suppose that we find a set of $|\tilde G|$ independent string operators,
$\{S_i\}$, obeying $[S_i,K_n]=0$ for all $i,n$ and $[S_i,S_j]=0$ for
all $i,j$.  We can repeat the argument in the lemma for each member of
the set, so, the rank of
\begin{equation}
\rho''=\prod_i(\identity+S_i)\prod_j(\identity+K_j)	\label{eqn:stringged}
\end{equation}
is 1, and hence no quantum bits can be stored in the degenerate ground
space of $H$. (This argument does not rule out storing up to $|\tilde G|$
classical bits.)

\subsection{General Construction of String Operators} \label{sec:proof}

In this section, we will describe how to construct two operators
$S_i^H$ and $S_i^V$, for each identity set of stabilizers, $G_i\in
G$. We show that, for each $G_i$, either
\begin{enumerate}
\item at least one of the
operators performs a non-trivial operation on the ground states, or
\item the set of stabilizers $G_i$ is defined over an area of fixed size, and cannot provide any protection against noise.
\end{enumerate}

We first introduce some notation that will be used throughout this
section.  Let us use $\bigwedge_AK$ to denote the restriction
of the operator $K$ to a particular area of the lattice $A$, replacing
all terms outside this area with $\identity$. So, for example, if the
area $A$ is just a single site, $\bigwedge_AK$ is just the Pauli operator
of $K$ that acts on that site.
\begin{definition}
  A set of stabilizers, $G_i^A$, is a subset of $G_i$ for which the
  elements $K_j$ satisfy
$$
\bigwedge_{\bar A}K_j=\identity,
$$
where ${\bar A}$ is the entire lattice not contained within the area
$A$.  In other words, the action of operators in $G_i^A$ is entirely
within $A$.
\end{definition}
We use $L_j^l$ to denote the area corresponding to a horizontal band
of height $j$, whose top edge coincides with row $l$ (according to a
numbering system of the rows with numbers increasing going down, and
counting is performed modulo $N$ to account for the periodic boundary
conditions).  Our constructions will involve horizontal strips of
spins of height $k-1$, across the whole width of the lattice
i.e.~$L_{k-1}^l$ for some $l$ \comment{($k$ will be the height of the largest
stabilizer in the particular identity set we are applying the
construction to **should probably rename it $\kappa$ to avoid
confusion with the original $k$**) I don't see the need for this statement, as there should never be a problem taking a strip height that's too large.}.  To simplify the notation, we
denote $L_{k-1}^l\equiv L, L_{2k-2}^l\equiv L_{\downarrow}$ and
$L_{2k-2}^{l-k+1}\equiv L_{\uparrow}$.  $\bigwedge_{L_{\downarrow}}$
selects all operators that are either entirely contained within $L$,
or extend below $L$, still overlapping with it. By taking the height
to be $k-1$, there is never a stabilizer which extends both above and
below $L$ (assuming a suitably large lattice size).  There are
equivalent terms involving vertical bands, but since horizontal and
vertical are arbitrary labels, constructions for both work
identically, although, importantly, they can give rise to
topologically inequivalent loops.

If we apply $\bigwedge_{L_m^l}$ to Eqn.~(\ref{eqn:degen}), we obtain
the identity
\begin{widetext}
\begin{equation}
  \bigwedge_{L_m^l}\left(\prod_{K\in G_i^{L_{m+k-1}^{l-k+1}}}K\right)\left(\prod_{K\in G_i^{L_m^l}}K\right)\bigwedge_{L_m^l}\left(\prod_{K\in G_i^{L_{m+k-1}^l}}K\right)=\identity,	\label{eqn:identity}
\end{equation}
which simply states that if we take the product of all stabilizers
that are not $\identity$ on a particular area\comment{qubit}, then the
overall product must be $\identity$ so that Eqn.~(\ref{eqn:degen}) is
satisfied.  We have chosen to state this for the area $L_m^l$, as this
will be most useful to us. There is a more general form of the
identity,
\begin{equation}
  \bigwedge_{L_m^l}\left(\prod_{K\in G_i^{L^{l-k+1}_r}}K\right)\bigwedge_{L_m^l}\left(\prod_{K\in G_i^{L_{r+l-k+1-n}^{n}}}K\right)\bigwedge_{L_m^l}\left(\prod_{K\in G_i^{L_{m+l+k-n-1}^n}}K\right)=\identity,	 \label{eqn:genident}
\end{equation}
\end{widetext}
which holds for any integers $n<l+k+m-1$, $r>n-l+k-1$.  Identity
(\ref{eqn:identity}) is the case $n=l$, $r=m+k-1$.  It is worth noting
that the strips $L$ for the first two terms have the same bottom edge,
and that the last two terms have the same top edge. Also, the top edge
of the first term and the bottom edge of the last term are such that
all possible operators that hit the strip $L_m^l$ are contained within
it.

We proceed to prove several important properties of the string
operators, that will be needed in the proof of Theorem
\ref{thm:main}.  The following identities are also useful:
\begin{equation}
K\left(\bigwedge_AK\right)=\bigwedge_{\bar A}K	\label{eqn:wedges}
\end{equation}
and
\begin{equation}
\left[\bigwedge_AK_1,K_2\right]=\bigwedge_A\left[K_1,K_2\right]\bigwedge_{\bar{A}}K_2,\label{eqn:commutators}
\end{equation}
where the latter identity follows from
\begin{equation*}
\bigwedge_A\left(K_1K_2\right)=\bigwedge_AK_1\bigwedge_AK_2.
\end{equation*}
and the definition of the commutator.

\begin{lemma} The operator $S_i^H$, defined by
\begin{equation}
S_i^H:=\bigwedge_{L}\left(\prod_{K\in G_i^{L_{\downarrow}}}K\right) \label{eqn:defS},
\end{equation}
satisfies $[S_i^H,K_n]=0$ for all $i,n$. \label{lem:proof1}
\end{lemma}
\begin{proof}
Identity (\ref{eqn:commutators}) gives
\begin{equation*}
[S_i^H,K_n]=\bigwedge_L\left[\prod_{K\in G_i^{L_{\downarrow}}}K,K_n\right]\bigwedge_{\bar{L}}K_n,
\end{equation*}
which equals zero because all the stabilizers commute.
\end{proof}
\begin{figure*}
\begin{center}
\includegraphics[width=0.9\textwidth]{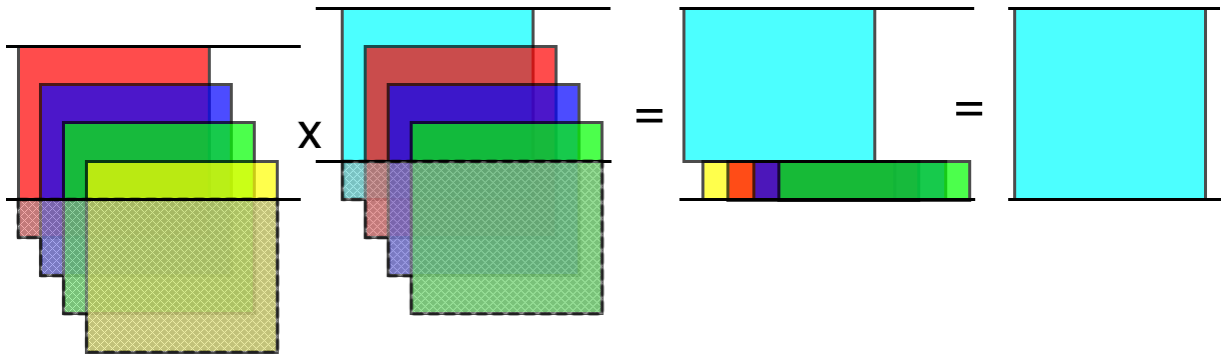}
\end{center}
\caption{To construct $S_i$ from $k\times k$ stabilizers, we consider a strip $L$ of height $k-1$ (the solid black lines). In the depicted example, $k=5$. Within this, we take all stabilizers that extend below the bottom line (these are the members of $L_{\downarrow}$), but truncate them to only include the parts in the area (so we remove the hatched components, acting $\bigwedge_L$ on the stabilizers). In this figure, we only depict one column of stabilizers, and offset them horizontally for clarity. If we were to construct the $S_i$ on one row higher, then by taking their product, we can see that they are related by the product of stabilizers. The final equality only holds when we consider all stabilizers along the strip, not just a single column.} \label{fig:construction}
\end{figure*}

Thus, for each set, $G_i$, we have found an operator that commutes
with all the stabilizers.  As we will show later, it is also
independent of $\identity$ and hence has non-trivial action on the
ground states (cf.\ Lemma (\ref{lem:action})).  Since $L$ is defined
with a specific upper row, $l$, one might apply this construction for
each $l$ (with fixed $G_i$) to generate further operators.  However,
such operators will not be independent as the following lemma shows.
\begin{lemma}
  Consider the operator $S_i^l$, a string operator $S_i^H$ where the
  corresponding area $L$ has its top edge on row $l$.  Then, for any
  $l$, $l'\neq l$, $S_i^l$ and $S_i^{l'}$ are dependent.  Further, the
  set under which they are dependent with respect to is $G_i$.
  \label{lemma:5}
\end{lemma}
\begin{proof}
  It suffices to show that $S_i^l$ and $S_i^{l-1}$ are dependent; the
  general relation then follows by induction. We have
  \begin{widetext}
\begin{equation*}
  S_i^lS_i^{l-1}=\bigwedge_{L_{k-1}^{l-1}}\left(\prod_{K\in
      G_i^{L^{l-1}_k}}K\right)\bigwedge_{L_{k-1}^{l-1}}\left(\prod_{K\in G_i^{L}}K\right)\bigwedge_{L_{1}^{l+k-2}}\left(\prod_{K\in G_i^{L_{\downarrow}}}K\right).
\end{equation*}
Using (\ref{eqn:wedges}) for $K\in G_i^L$ gives
$\bigwedge_{L^{l-1}_{k-1}}K=K\bigwedge_{L_1^{l+k-2}}K$ etc.,
hence
\begin{eqnarray*}
  S_i^lS_i^{l-1}&=&\left(\prod_{K\in G_i^{L^{l-1}_k}}K\right)\left(\prod_{K\in G_i^{L}}K\right)\bigwedge_{L_{1}^{l+k-2}}\left(\prod_{K\in G_i^{L^{l-1}_k}}K\right)\bigwedge_{L_{1}^{l+k-2}}\left(\prod_{K\in G_i^{L}}K\right)\bigwedge_{L_{1}^{l+k-2}}\left(\prod_{K\in G_i^{L_{\downarrow}}}K\right)\\
  &=&\left(\prod_{K\in G_i^{L^{l-1}_k}}K\right)\left(\prod_{K\in G_i^{L}}K\right)
\end{eqnarray*}
\end{widetext}
where the final line follows from (\ref{eqn:genident}).
\end{proof}
Thus, the operators $S_i^l$ and $S_i^{l'}$ $l'\neq l$ are dependent
and hence Lemma \ref{lem:sameact} implies that they have the same
action on the ground states.  Similarly, all column operators for a
particular $G_i$ are related by products of stabilizers. Provided the
row and column operators implement closed loops around the boundaries
of the lattice (i.e.\ around inequivalent loops of a torus), they are
topologically inequivalent, and therefore cannot be related by
products of stabilizers. \comment{However, if the sets do not involve
  such loops, the row and column operators $S_i^H$ and $S_i^V$ are
  dependent with respect to $G_i$.}

\begin{lemma}
  If the operator $S_i^H$ formed from a set $G_i$ is dependent on
  $\identity$, then it can be written as a product of stabilizers
  entirely contained within $L$.
  \label{lem:proof2}
\end{lemma}
\begin{proof}
  Suppose that $S_i^H$ is dependent on $\identity$, i.e.\ that it can
  be written in the form $\prod_{K\in W}K$ for some set $W$.  In order
  that $S_i^H$ be $\identity$ on $\bar{L}$, $W$ must take all the
  members of sets $G_k\setminus G_k^L$, for some $k$ or none of them.
  To see that this is a necessary condition, note that $\prod_{K_j\in
    U}K_j=\identity$, with $U^L=\emptyset$ implies $U=G_k$ for some
  $k$.  Furthermore, since $\prod_{K\in G_k^L}K=\prod_{K\in
    G_k\setminus G_k^L}K$ for all $k$, we can always redefine $W$ as
  $W'$ such that $W'=W'^L$.
\end{proof}

\begin{lemma}
  \label{lemma:7} An elementary set $\tilde G_i$ that is topologically
  trivial horizontally (vertically) can be replaced with an elementary
  set consisting of stabilizers defined within a region of width $k$
  (height $k$).
\end{lemma}
\begin{proof}
  Lemma \ref{lemma:5} tells us that the operators formed by using our
  construction on the same identity set, but starting in different
  places are dependent.  Since the set $\tilde G_i$ is topologically
  trivial, we can define a top row, the $t$th row, the highest row
  which contains a stabilizer.  Consider forming the horizontal string
  operator starting at the top (i.e.\ forming $S^t_i$ in the notation
  of Lemma \ref{lemma:5}).  We have
  $S^t_i=\bigwedge_{L^{t}_{k-1}}\left(\prod_{K\in\tilde{G}_i}K\right)=\identity$,
  since when choosing this top row, we can replace
  $\tilde{G}_i^{L_{\downarrow}}$ by $\tilde{G}_i$.  Hence $S_i^H$ must
  be dependent on $\identity$, for an arbitrary starting row.

  Consider then constructing the horizontal string operator $S^{t+1}_i$.  This operator is
  dependent on identity, so is a product of stabilizers.  Furthermore,
  it is entirely contained in a strip of height $k-1$.  From the
  argument in Lemma \ref{lem:proof2}, we can define it in terms of a
  set of stabilizers contained in this strip.  We write
  $S^{t+1}_i=\prod_{K\in W}K$, where $W$ is a set of stabilizers
  entirely within the strip.

  Now, consider the stabilizers from $\tilde{G}_i$ whose top row is
  $t$.  Denote this set by $T$.  We can write
  $S^{t+1}_i=\bigwedge_{L^{t+1}_{k-1}}\left(\prod_{K\in\tilde{G}_i\setminus
        T}K\right)$ (recall that $\bigwedge_{L^{t+1}_{k-1}}$ simply
    removes the area below row $t+k$).  We have
    $\prod_{K\in\tilde{G}_i\setminus T}K\prod_{K\in T}K=\identity$.
    Thus,
  $$
  \bigwedge_{L^{t+1}_{k-1}}\left(\prod_{K\in\tilde{G}_i\setminus
      T}K\right)\bigwedge_{L^{t+1}_{k-1}}\left(\prod_{K\in T}K\right)=\identity.
  $$
  However, $\bigwedge_{L^{t+1}_{k-1}}\left(\prod_{K\in
      T}K\right)=\prod_{K\in T}K$, hence $\prod_{K\in W}K\prod_{K\in
    T}K=\identity$.  We have thus found a new identity set contained
  entirely within a strip of height $k$.
\end{proof}

\begin{corollary}
  \label{cor:1}
  An elementary set $\tilde G_i$ that is topologically trivial in both
  directions can be replaced with an elementary set consisting of
  stabilizers defined within a $k\times k$ area.
\end{corollary}

We have hence shown that the elementary sets can be chosen such that
each is either local or topologically non-trivial.  It remains to
prove that for the subsets $\tilde G_i$, the operators break the
corresponding degeneracy and are independent of each other.

First we argue that for a local elementary set, $\tilde{G}_i$, there
exists a local operator that will break the degeneracy.  To see this,
consider a $k\times k$ lattice with a Hamiltonian containing only the
stabilizers in $\tilde{G}_i$.  There are $M_k=2^{k^2-|\tilde{G}_i|+1}$
groundstates, hence $M_k$ commuting operators on this $k\times k$
region breaking these degeneracies.  If one then expands the lattice
to be $N\times N$ (without adding any new stabilizers), then this same
set of localized operators splits the degeneracy.  Now consider adding
back the other stabilizers.  They break many of the $M_k$ degeneracies,
but by assumption, one remains.  However, all of the operators
breaking such a degeneracy are local.

It is therefore impossible to get self-correction by encoding
information in grounds states whose degeneracy is caused by a local
elementary set.

\begin{lemma}
  \label{lemma:8}
  The operator $S_i^H$ formed from a set $G_i$ that is composed of
  only topologically non-trivial elementary sets in the vertical
  direction is independent of $\identity$.
\end{lemma}
\begin{proof}
  Consider a set of stabilizers $W_0$ that is the union of $W_R$ and
  $G_i$, so that the only identity sets that are present are due to
  $G_i$. Now assume that, contrary to the lemma, $S_i^l=\prod_{K\in
    W_l}K$. From Lemma \ref{lem:proof2}, we can take $W_l$ such that
  $W_l=W_l^L$. This may require the enlargement of set $W_0$, which
  could introduce new identity sets.  However, any identity sets that
  are introduced cannot be restricted to the area $L$ (otherwise it
  would not be necessary to introduce the set), and hence must be topologically non-trivial in the vertical direction. Now, recall from Lemma
  \ref{lemma:5} that $S_i^{l-1}S_i^l=\prod_{K\in W'}K$ where
  $W'\subseteq G_i$.  We have
  \begin{equation}
    \prod_{K\in W_l}K\prod_{K\in W_{l-1}}K\prod_{K\in W'}K=\identity.
  \end{equation}
  So, if $(W_l,W_{l-1})\nsubseteq G_i$, we have formed a new identity set
  from members of $W_l, W_{l-1}$ and $W'$, in contradiction with the
  assumption.

  The remaining possibility is that $(W_l,W_{l-1})\subseteq G_i$.  One
  way to satisfy this is if $W'$ has members not in $(W_l,W_{l-1})$.
  Then we have formed an identity set which is topologically trivial
  vertically, a contradiction.  Alternatively, all members of $W'$ are
  also members of $(W_l,W_{l-1})$, so do not have height more than
  $k-1$, instead of the $k$ that we were assuming. Thus, one reapplies the construction of $S^H_i$ such that it has height $k-2$. Either one of the previous cases occurs, and thus the lemma holds, or we conclude that the stabilizers must have height $k-2$. Hence, we can continue making this same argument until we either conclude that the lemma holds, or the height of the stabilizers is 0, i.e.~$G_i=\emptyset$.
\end{proof}

We state a corollary of this lemma which follows because the product
$S^H_iS^H_j$ is equal to the corresponding operator of the combined
set $(\tilde{G}_i,\tilde{G}_j)$.
\begin{corollary}
\label{lemma:9}
  For any pair of elementary sets, $\tilde{G}_i$ and $\tilde{G}_j$,
  which are topologically non-trivial in the vertical direction, the
  corresponding operators, $S^H_i$ and $S^H_j$ are independent.
\end{corollary}

Note that if $\tilde{G}_i$ is topologically non-trivial vertically,
but topologically trivial horizontally, then (cf.\ Lemma
\ref{lemma:7}) one could redefine $\tilde{G}_i$ such that it contained
stabilizers in a strip of width $k$.  The $S_i^H$ formed from this set
will therefore be local.  It follows that we only generate a loop
operator for elementary sets which are topologically non-trivial in
both directions.

\begin{lemma}
  \label{lemma:10}
  Let $\tilde{G}_i$ be an elementary set which is topologically
  non-trivial in the vertical direction, but topologically trivial
  horizontally, and $\tilde{G}_j$ be an elementary set which is
  topologically non-trivial in the horizontal direction, but
  topologically trivial vertically.  The corresponding operators,
  $S^H_i$ and $S^V_j$, formed from these sets are independent.
\end{lemma}
\begin{proof}
  Since $\tilde{G}_i$ and $\tilde{G}_j$ are topologically trivial in
  one direction, we can use Lemma \ref{lemma:7} to redefine them as
  having a strip of width, respectively height, $k$.  The
  sets then have overlap in a $k\times k$ area.  We choose this area
  to form $S_i^H$ and $S_j^V$. Moreover, the operators
  $S_i^{l-1}S_i^l$ and $S_j^{m-1}S_j^m$ are also restricted to this
  $k\times k$ block (with upper-left corner $(m-1,l-1)$). We can now
  make an argument that closely parallels that of Lemma \ref{lemma:8}.

  As in Lemma \ref{lemma:8}, we form a set $W_0$, here as the union of
  $W_R$, $\tilde{G}_i$ and $\tilde{G}_j$.  Assume that $S_i^H$ and
  $S_j^V$ are dependent, i.e.~$S_i^lS_j^m=\prod_{K\in W_{l,m}}K$,
  where each of the members of $W_{l,m}$ must have either height or
  width $k-1$. As before, we must have
  \begin{equation}
    \prod_{K\in W_{l,m}}K\prod_{K\in W_{l-1,m}}K\prod_{K\in W_{l,m-1}}K\prod_{K\in W'}K=\identity.
  \end{equation}
  where $W'\subseteq (\tilde G_i,\tilde G_j)$. If
  $(W_{l,m},W_{l-1,m},W_{l,m-1})$ contains elements not in $(\tilde
  G_i,\tilde G_j)$, then there exists a topologically trivial identity
  set, a contradiction. The same holds if $W'$ contains elements not
  in $(W_{l,m},W_{l-1,m},W_{l,m-1})$, which means that no members of
  $\tilde G_i$ can have height $k$, and no members if $\tilde G_j$ can
  have width $k$, unless $\tilde G_i$ and $\tilde G_j$ contain the
  same terms of size $k\times k$. It is not possible for $\tilde G_i$
  and $\tilde G_j$ to contain the same $k\times k$ term, since, were
  we to remove it, it would break both identity sets, which is in
  contradiction with the definition of elementary sets. Thus we can
  repeat this construction for strips of dimension $k-2$, and the
  argument recurses (in the same was as in Lemma \ref{lemma:8}) until
  we conclude that the lemma holds, or the sets are null.
\end{proof}

We are now ready to prove our main theorem.
\begin{proof}[Proof of Theorem \ref{thm:main}]
  For every elementary set $\tilde{G}_i$, there are two possible forms
  for the logical operations.  If $\tilde{G}_i$ is topologically
  trivial, then there is a local operator breaking the relevant
  degeneracy.  Any data encoded in ground states that are degenerate
  due to such a set is not protected against thermal noise or
  Hamiltonian perturbations (due to its fixed size).

  If $\tilde{G}_i$ is topologically non-trivial vertically, then we
  can generate a string operator $S_i^H$ breaking the degeneracy, and
  likewise if it is topologically non-trivial horizontally, then we
  can generate $S_i^V$ breaking the degeneracy.  It follows from
  Corollary \ref{lemma:9} and Lemma \ref{lemma:10} that for different elementary sets, these are
  independent.  We have therefore found a way to break all the
  degeneracies caused by products of stabilizers being identity.

  To complete the proof of Theorem \ref{thm:main}, it remains to show
  that any string operators $S_i^H$ and $S_i^V$ provide error paths for
  thermal noise.  If the operators are not point-like, then they take the form of a
  loop of fixed width around the torus.  If we truncate such a loop,
  then only a finite number ($O(k^2)$) of stabilizers do not commute
  with it, so there is an approximately constant energy cost as the
  amount of truncation is varied. The ends of the truncated string
  thus behave precisely like the strings in the toric code. Hence, the
  survival time against thermal noise is not exponential, and these
  codes are not self-correcting.
\end{proof}

It may also serve as a useful side observation that the truncations of
our string operators give the excited states of the Hamiltonian,
describing pairs of anyons. Since the strings are tensor products of
Pauli operators, two such strings either commute or anti-commute, and
Abelian braiding properties of the anyons can be realised, which are
identical to those of the toric code, except with up to $|\tilde G|$
different particle types, and subsequent composite particles.

\section{Higher Spatial Dimensions}

While our constructions have been formed specifically for 2D square
lattices, they work equally well on arbitrary 2D geometries. We can
also choose to apply an identical construction in $d$ dimensional
systems to find the structure of one logical operation per qubit. The
proof proceeds as before, replacing the loop $L$ with a
$(d-1)$-dimensional object with height $k-1$ in the other dimension.

With $d=3$, for example, we arrive at an area operator, which provides
protection for classical data. For each elementary set, we
take one such operator, all of which act in the same plane, to define
the $Z$-basis of our qubits. What happens for the other logical gate
operation? Clearly, to generate the $SU(2)$ algebra, we would need an
operation that anticommutes with the plane (we can shift the
plane to arbitrary positions).  Thus, the only logical structure is a
string-like object, although we have no explicit construction for its
form.  This argument suggests that self-correction is also impossible
for 3D stabilizer Hamiltonians.

On the other hand, in $d=4$, the same argument does not apply. This is
because although our construction can give a 3D object, if the
stabilizers in an identity set only span 3 dimensions themselves, the
logical operation is two-dimensional, and an anti-commuting
2-dimensional term can also be found, giving protection to a whole
qubit, which is identical to the classical protection afforded by the
2D Ising model. This is the case in the 4D toric code.

\section{Conclusions}

In this paper, we have shown that for all qubit stabilizer
Hamiltonians in 2D with periodic boundary conditions (subject to the condition that the Hamiltonian is
specified), there are always string operators that loop around the
torus, and perform non-trivial actions on the ground state space.
There are always enough of these to ensure that there is no
non-trivial subspace of the ground state space which is not affected,
and hence there is no possibility of storing quantum data that cannot
be affected by such an operator.  Furthermore, we give an explicit
method for constructing the operators that give the paths for noise.
Given the existence of these, we are led to conclude that none of
these systems are self-correcting, in the same way that the toric
code is not self-correcting.  In addition, we suggest that a similar
construction might work for stabilizer Hamiltonians in 3D.

There are several potential routes for further investigation. It would
be interesting to understand if this proof for stabilizer Hamiltonians
can be applied to non-stabilizer Hamiltonians. In particular, many
Hamiltonians are in some sense `close' to a stabilizer Hamiltonian, so
it might seem surprising if these could exhibit self-correction.
However, there are radically different Hamiltonians with, for example,
chiral terms \cite{pachos-2004-70}, which stabilizer codes cannot
encapsulate, and clearly it would be interesting to investigate if
these can be self-correcting.\bigskip

{\bf Additional Note:} We recently became aware of independent work of
a similar nature \cite{terhal:08}.

\acknowledgments ASK is supported by DFG (FOR 635 and SFB 631), EU
(SCALA), and Clare College, Cambridge, and would like to thank F.~Verstraete and I.~Cirac for useful discussions.  RC thanks Homerton
College, Cambridge for financial support during the initial stages of
this work.

\end{document}